# Addressing spatial dependence in technical efficiency estimation: A Spatial DEA frontier approach

Julián Ramajo[a] ✉ , Miguel A. Márquez[a] , Geoffrey J.D. Hewings [b]

[a] Department of Economics, University of Extremadura

Avenida de Elvas, s/n – Badajoz, E-06006 (Spain)

[b] Regional Economics Applications Laboratory, University of Illinois

1301 W Gregory Drive, Suite 65-67 – Urbana, IL 61801 (USA)

**E-mail addresses:**

ramajo@unex.es; mmarquez@unex.es; hewings@illinois.edu

**ORCiDs:**

0000-0002-3156-8315; 0000-0003-0784-5025; 0000-0003-2560-3273

**Funding details**:

This work was supported by the Ministry of Science and Innovation of Spain under Grant PID2019-109687GB-I00.

# Addressing spatial dependence in technical efficiency estimation: A Spatial DEA frontier approach

**Abstract**

This paper introduces a new specification for the nonparametric production-frontier based on Data Envelopment Analysis (DEA) when dealing with decision-making units whose economic performances are correlated with thoses of the neighbors (spatial dependence). To illustrate the bias reduction that the SpDEA provides with respect to standard DEA methods, an analysis of the regional production frontiers for the NUTS-2 European regions during the period 2000-2014 was carried out. The estimated SpDEA scores show a bimodal distribution do not detected by the standard DEA estimates. The results confirm the crucial role of space, offering important new insights on both the causes of regional disparities in labour productivity and the observed polarization of the European distribution of per capita income.

*Keywords:* Production frontier, Data envelopment analysis, Conditional efficiency, Geographical externalities, European regions.

## 1   Introduction

In their introduction to a collection of articles addressing the Rebuilding Macroeconomic Theory Project, Vines and Wills (2018) noted that four main changes to the core model were recommended. These were the needs to (1) emphasize financial frictions; (2) place a limit on the operation of rational expectations; (3) devise more appropriate microfoundations and (4) include heterogenous agents. While the latter concern was addressed in terms of heterogenous consumption behavior of households of different size or income levels, none of the authors considered the role and impact of heterogenous space. No attention was directed to the problem of differences in economic structure and performance at the sub-national (regional) levels and the way in which ignoring this spatial heterogeneity might create problems for the effective performance of macroeconomic models. Further as Proost and Thisse (2019) have noted, there is an important link between microeconomic spatial forces/foundations and their contributions to generating and sustaining spatial inequalities. The role and importance of spatial spillovers associated with flows of goods and services, people, and innovations create dynamics that can have profound implications for the performance of the macroeconomy.

This paper explores the role of the spatial dimension in the context of concerns about the increase of global income inequality (Alvaredo et al., 2018) especially the absence of income convergence between countries (Johnson and Papageorgiou, 2020) and the rising imbalances within countries (Blanchet et al., 2019). These general trends are of special relevance for the case of the European Union (EU), where increasing regional labour productivity is a major challenge for regions that aim to raise their incomes because the lagging regions in Europe are characterized by low productivity levels (Widuto, 2019). Recently, underlying and interdependent spatial factors have been identified as fundamental drivers of the productivity dynamics (Tsvetkova et al., 2020). Moreover, regional disparities in the EU could be related to the important issue of the productivity gap in Europe (Bruno et al., 2019). Therefore, a diagnosis of the determinants of the ongoing divergences in the EU should provide a measure of the performance of the regional economies in Europe.

In this framework, the technical efficiency of a regional economy would be the degree to which its production level approaches the maximum output obtainable with its disposable inputs (Färe and Lovell, 1978). In recent decades, applied researchers have become increasingly interested in the analysis of efficiency across regional economies in order to determine whether or not the use of regional factor endowments is contributing to regional convergence (Maudos et al., 2000;

Ramajo and Hewings, 2018). This interest is related to the important role that changes in efficiency could have on changes in productivity (Solow, 1957) affecting the regional convergence process. One strand of this literature has focused on the analysis of regional technical efficiency (Fritsch and Slavtchev, 2011; Tsekeris and Papaioannou, 2018). Regions face different contextual variables related to the location in which they operate and some of these observed heterogeneous conditions (environmental factors), being beyond the control of local authorities, affect their production process. Consequently, the estimation and assessment of efficiency in regional production should consider the spatial dimension of the decision-making units (DMUs) analyzed. Spatial dependence in regional production efficiency refers to the correlation between the economic performance of the regions and those of the neighboring regions. Nevertheless, the majority of the existing empirical studies have failed to consider the existence of spatial dependence in the estimation of the regional efficiency frontier. For example, Aiello and Bonanno (2019) analyze the differences in efficiency in the local government literature by reviewing 54 papers published during the period 1993-2016 and, in spite of the recognition of the relevance of the space in this type of studies, the review does not contemplate the possible existence of spatial factors when estimating local efficiencies.

To measure technical efficiency, one of the most popular nonparametric approaches is Data Envelopment Analysis (DEA), introduced by Charnes et al. (1978, 1981). Focusing on this enterely nonparametric technique (Van Biesebroeck, 2008), it is surprising that it has not included any spatial element in its formulation (see, for example, Emrouznejad and Yang, 2018, or Ray, 2020). This lack of spatial consideration could be causing biases in the estimation of the technical efficiency scores, hiding relevant information on the performance of the analyzed DMUs and providing unsatisfactory results in the benchmarking analysis of units' performance. This paper intends to fill this literature void, by providing a new contribution to the production economics literature, the Spatial DEA (SpDEA) model, which can shed light to assess the efficiency of spatially referenced DMUs. We apply our proposed SpDEA specification to 263 European NUTS-2 regions over the period 2000-2014. It was found that the results of the two models (DEA and SpDEA) are different. In fact, after controlling for geographical externalities, the average regional efficiency increased with respect to the standard DEA results. Other important finding is that the shape of the densities associated with the efficiency scores estimated by the SpDEA model for each year (2000, 2007, and 2014) show a tendency towards a bimodal or twin-peaked distribution; this twin-peakedness is not detected in the DEA efficiency scores. The relevance of these results comes from the fact that they are coherent with the empirical analyses carried out previously in terms of regional productivity and convergence for the case of world countries (Henderson and Russell, 2005; Badunenko et al., 2013, 2018) or the European regions (Rogge, 2019), shedding light on the causes of the observed regional performance inside the EU. However, this bimodal regional efficiency in Europe would be hidden if the biased scores from the standard (aspatial) DEA were used.

The rest of the paper is organized as follows. The methodological framework of the analysis and the proposed SpDEA approach are presented in section 2. The data and variables are described in section 3. Also, in this empirical part, we estimate the standard DEA and the SpDEA model to account for regional spillovers. Section 4 summarizes the results and offers some conclusions.

## 2   Space-dependent efficiency scores: the SpDEA model

This section starts by placing the contribution of this work in context showing recent developments in the nonparametric frontier literature. Later, we overcome the limitation detected in the standard DEA technique when dealing with georeferenced DMUs by means of the proposed SpDEA model.

**The standard (output-oriented) DEA method**

Following the seminal work of Färe et al. (1994), this paper addresses the issue of the relative technical efficiency by means of the DEA approach, which does not require the specification of the functional form of the production function being estimated. In order to measure the efficiency of regions, we define a set of $p$ inputs $\boldsymbol{x} = (x_1, x_2, \ldots, x_p) \in \mathbb{R}_+^p$ that are used to produce a vector of $r$ outputs $\boldsymbol{q} = (q_1, q_2, \ldots, q_r) \in \mathbb{R}_+^r$. Then the technology set of all feasible input-output combinations $(\boldsymbol{x}, \boldsymbol{q})$ can be defined as:

$$\psi = \{(\boldsymbol{x}, \boldsymbol{q}) \in \mathbb{R}_+^{p+r} \mid \boldsymbol{x} \text{ can produce } \boldsymbol{q}\} \tag{1}$$

The unconditional output-oriented Farrell-Debreu technical efficiency DEA score, $\hat{\lambda}_{DEA}$, of a DMU can be obtained by solving the following linear programming problem:

$$\hat{\lambda}_{DEA}(\boldsymbol{x}, \boldsymbol{q}) = sup\{\lambda > 0 \mid (\boldsymbol{x}, \lambda \boldsymbol{q}) \in \hat{\psi}\} \tag{2}$$

where $\hat{\psi} = \{(\boldsymbol{x}, \boldsymbol{q}) \in \mathbb{R}_+^{p+r} \mid \boldsymbol{q} \leq \sum_{i=1}^n \gamma_i \boldsymbol{q}_i, \boldsymbol{x} \geq \sum_{i=1}^n \gamma_i \boldsymbol{x}_i, \sum_{i=1}^n \gamma_i = 1, \gamma_i \geq 0\}$ is the attainable set estimated from an observed random sample of DMUs $\{(\boldsymbol{x}_i, \boldsymbol{q}_i) \mid i = 1, 2, \ldots, n\}$, and $\lambda$ is the efficiency parameter to be evaluated for the productive unit operating at level $(\boldsymbol{x}, \boldsymbol{q})$. Using this definition, $\hat{\lambda}_{DEA}(\boldsymbol{x}, \boldsymbol{q}) = 1$ denotes an efficient production unit, while $\hat{\lambda}_{DEA}(\boldsymbol{x}, \boldsymbol{q})^{-1} < 1$ implies that the corresponding DMU is inefficient.

**The influence of contextual variables and conditional DEA estimator**

To control for spatial dependence generated by geographically referenced data, our DEA specification will incorporate spatial autoregressive terms as external conditioning factors in order to estimate regional efficiency scores, using the conditional nonparametric approach proposed by Daraio and Simar (2005, 2007). As a result, we can explore the potential influence of interregional spillovers in the regional production process but, contrary to the parametric production approach (where the spatial effects are associated with specific parameters measuring the relevance of geographical externalities across cross-sectional units), in our spatialized DEA model the effects of the spatial factors are not just one-directional (positive or negative) but they can have a variable influence on the productivity performance of regions.[1]

To analyze the influence of contextual variables (in our case spatial external factors) on the production function or the inefficiency distribution, we need to reformulate the optimization problem associated with the baseline DEA model. In this case, the production process must be defined by using the alternative probabilistic formulation following the notation introduced by Cazals et al. (2002), and Daraio and Simar (2005, 2007). In this approach, the production process can be described by the probability measure of the joint random vector $(X, Q)$ denoted by $H_{X,Q}(x, q)$, which represents the probability of dominating a unit operating at level $(x, q)$:

$$H_{X,Q}(x, q) = \text{Prob}(X \leq x, Q \geq q) \quad (3)$$

This probability function can be further decomposed as follows:

$$H_{X,Q}(x, q) = \text{Prob}(Q \geq q | X \leq x)\text{Prob}(X \leq x) = S_{Q|X}(q|X \leq x)F_X(x) \quad (4)$$

where $S_{Q|X}(q|X \leq x)$ represents the conditional survival function of $Q$ and $F_X(x)$ is the cumulative distribution function of $X$.

The probabilistic formulation of the production process is able to handle the presence of observed heterogeneity in the form of contextual $k$ factors $z = (z_1, z_2, \ldots, z_k) \in \mathbb{R}_+^k$ that might have an influence on the production process. For this purpose, a conditional efficiency approach can be used to account for such variables in the frontier estimation by conditioning the production process to a given value of the environmental random vector $Z$, $Z = z$.

---

[1] Chung and Hewings (2015) have shown in the context of economic interdependence associated with business cycle behavior that regions (in this case, US states) can display asymmetric relationships with each other.

Consequently, efficiency estimates are determined by inputs, outputs and external variables. The joint probability function can be extended and decomposed as follows:

$$H_{X,Q|Z}(x, q|z) = \Pr(Q \geq q | X \leq x, Z = z) \Pr(X \leq x, Z = z) =$$

$$= S_{Q|X,Z}(q | X \leq x, Z = z) F_{X|Z}(x | Z = z) \quad (5)$$

The function $H_{X,Q|Z}(x, q|z)$ represents the probability of a DMU operating at level $(x, q)$ being dominated by other DMUs facing the environmental conditions $z$ that might influence the production process but not controlled by the production units.

Using this formulation, the conditional DEA (cDEA) output measure of technical efficiency can be obtained as:

$$\hat{\lambda}_{cDEA}(x, q|z) = sup\{\lambda > 0 \mid \hat{S}_{Q|X,Z}(\lambda q | X \leq x, Z = z) > 0\} \quad (6)$$

where $\hat{\lambda}_{DEA}(x, q)^{-1} \leq \hat{\lambda}_{cDEA}(x, q|z)^{-1} \leq 1$ since for all $z$, $\psi^z = \{(x, q) | Z = z, x \text{ can produce } q\} \subseteq \psi$.

The estimated function $\hat{S}_{Q|X,Z}$ in (6) is harder to evaluate than in the unconditional case (the support set $\hat{\psi}$ in equation (2)), because it requires the use of smoothing techniques for the external variables in $z$ (Bădin et al., 2010):

$$\hat{S}_{Q|X,Z}(q | X \leq x, Z = z) = \frac{\sum_{i=1}^{n} \mathbb{I}(x_i \leq x, q_i \geq q) K_h(\frac{z_i - z}{h})}{\sum_{i=1}^{n} \mathbb{I}(x_i \leq x) K_h(\frac{z_i - z}{h})} \quad (7)$$

where $\mathbb{I}(\cdot)$ is the indicator function. Therefore, this approach relies on the selection of a product kernel function $K_h(\cdot)$ with compact support and an optimal bandwidth parameter $h > 0$ selected using any choice method. In the empirical application of section 3, we adopt the data-driven selection approach developed by Bădin et al. (2010) that is based on the Least Squares Cross-Validation (LSCV) method (see the recent work of Bădin et al., 2019, where practical aspects of bandwidth selection for conditional efficiency are exposed).

By means of the conditional frontier approach, the direction of the effect of the external factors on the production process can be evaluated by comparing conditional with unconditional efficiency scores (Bădin et al., 2012). Using the ratio of efficiency estimates, it will be possible to analyze the impact that $Z$ may have on the shape of the efficient boundary (shifts in the frontier), in such a way that the conditioning variables can affects the range of attainable values of the input-output space.

More specifically, as will become clear in our empirical application, if the *z*'s are continuous variables, then the influence of the environmental variables on the distribution of the efficiency scores can be quantified by computing the graph of the ratios of conditional to unconditional scores (see expression 8) against the components of $z_i$ and its smoothed nonparametric regression function over the sample of *n* observations $\{(x_i, q_i, z_i) | i = 1,2, ..., n\}$ (Daraio and Simar, 2005, 2007):

$$\hat{R}(x_i, q_i | z_i) = \frac{\hat{\lambda}_{cDEA}(x_i, q_i | z_i)}{\hat{\lambda}_{DEA}(x_i, q_i)} \tag{8}$$

By generating these partial regression plots, we can easily inspect the direction of the effect of the environmental variables upon the efficiency ratios: in our output-oriented conditional model, an increasing regression curve indicates that the corresponding variable in $z$ is favorable to efficiency (the conditional frontier moves to the unconditional one when the concrete variable *z* increases), whereas a decreasing line will denote an negative effect on the production process (the conditional efficient boundary moves away from the unconditional frontier).

Likewise, it is also possible to investigate the statistical significance of the explanatory variables $z$ explaining the variations of $\hat{R}(x_i, q_i | z_i)$. Following Badin et al. (2010), and Jeong et al. (2010), local linear least squares can be used for the regression estimation, followed by the nonparametric regression significance test proposed by Li and Racine (2004). Specifically, the significance of each variable is tested using bootstrap resampling tests as proposed by Racine (1997) and Racine et al. (2006); this approach can be interpreted as the nonparametric equivalent of standard *t*-ratio tests in ordinary least squares regression.

**Spatialized frontier models and the new SpDEA specification**

Although it is well known that disregarding spatial aspects of the data may produce inefficient or even biased estimates of efficiency scores, only recently have there been studies that have tried to estimate technical efficiency incorporating information about the structure of the spatial heterogeneity and/or spatial dependence of geographically distributed productive units. In response, a new class of spatial frontier models has appeared in the productivity and efficiency analysis literature (see the recent survey of Orea and Álvarez, 2020, and the references therein). The main message that emerges from this piece of literature is the importance of considering uneven performance in space (spatial heterogeneity), geographical interconnexion (spatial dependence), and spillover effects. Then, the need to formulate specific models for spatially structured data can be concluded.

In the context of our empirical application, in order to build a spatial frontier model for the EU-28 regions, we begin postulating a regional production function $Y_{it} = F(K_{it}, L_{it})$, where $Y_{it}$ represents the aggregate output in region *i* at year t, $K_{it}$ denotes the respective stock of physical capital, and $L_{it}$ is the employed labor force. As is usual in the productivity growth literature at the macro level, the constant returns to scale (CRS) assumption will be used (at the national level see, for example, Färe et al., 1994, Kumar and Russell, 2002, Henderson and Russel, 2005, or Los and Timmer, 2005; at regional level see, among others, by Enflo and Hjertstrand, 2009, Filippetti and Peyrache, 2015, or Beugelsdijk et al., 2018). Hence, in the production frontier, the output per worker (labour productivity) ratio ($q_{it} = Y_{it}/L_{it}$) is a function only of the capital per worker ('capital deepening') input ($x_{it} = K_{it}/L_{it}$), $q_{it} = f(x_{it})$; in essence, this production model implies that for any given per worker input of physical capital there is an efficient level of output per worker. In other words, under this simple unconditional frontier approach the difference in the level of labour productivity between an efficient region and another inefficient one is due only to two components, the distance of the inefficient region to the frontier, and the different capital intensity (level of capital stock per unit of labour) between the two regions. However, this difference does not depend on the location of the compared regions.

To introduce the spatial dimension in the DEA model, accounting for the spatial dependence observed in the regional data, a spatial lag of the output per worker ratio (a weighted average of the level of productivity in nearbouring regions), $z_{1i} = \sum_{j=1}^{n} w_{ij} q_j = \boldsymbol{w}_i' \boldsymbol{q}$, and a spatial lag of the capital-labour ratio (a weighted average of capital intensity in nearby regions), $z_{2i} = \sum_{j=1}^{n} w_{ij} x_j = \boldsymbol{w}_i' \boldsymbol{x}$, are used as components of external conditioning factors of the vector $\boldsymbol{z}$ in the DEA efficiency approach.[2] The spatial weights $w_{ij}$ capture the spatial interaction between regions $i$ and $j$ (these elements are known a priori and satisfy the conditions $w_{ij} \geq 0$, $w_{ii} = 0$, and $\boldsymbol{w}_i' \boldsymbol{1} = \sum_{j=1}^{n} w_{ij} = 1$), and the resulting e $n$ by $n$ matrix $\mathbf{W}$ with elements $w_{ij}$ describes the connectivity of the $n$ regions.[3]

From a methodological point of view, our proposed spatialized DEA specification could be considered as the nonparametric counterpart to the Spatial Durbin Frontier model used in the parametric stochastic frontier literature, where a spatially lagged output and spatial lags of the inputs quantities are introduced in the stochastic frontier specification. So, our SpDEA nonparametric model is in essence similar to the parametric models proposed in Glass et al., 2016a, and Ramajo and Hewings, 2018, where a spatial autoregressive term (the $z_1$ variable in our formulation) to account for global spatial externalities (representing the spillovers of the effects of labour productivity in nearby regions) and also the spatial lags of the inputs (the $z_2$ variable in the conditional model) that account for local spillover effects (of capital intensity in our case) are introduced in the frontier specification.

---

[2] The inclusion of the *w'q* and *w'x* variables can be rationalized from a theoretical point of view if we think in a production function such as $Y = AF(K, L)$ where $A$ (which represents thecnological progress, incorporating aspects of knowledge generation, innovation, commuting, sectoral linkages or level of human capital at the NUTS-2 regional level) depends on the spatial lags of both endogenous and exogeneous variables. See, for example, Kock (2010), where a spatially augmented Solow parametric model with physical capital externalities and spatial externalities is developed and estimated using 204 European NUTS-2 regions over the 1977-2000 period.

[3] In our empirical application, we used a weight matrix defined as $\mathbf{W} = \{w_{ij} = 1 \text{ if } i \neq j \text{ and } j \in N_5(i); w_{ij} = 0 \text{ otherwise}\}$, $N_5(i)$ being the set of 5-nearest neighbors to $i$. We chose five neighbors as cut-off point to the connections of $i$ because 5 was the median of neighbors between the spatial units of our sample data. In addition to the 5-nearest neighbor matrix, we also used the traditional (first order) contiguity matrix and the inverse distance matrix as alternative weighting matrices, and the results were almost identical to the presented in section 4. These complementary results are disposable upon request to the corresponding author.

Figure 1 provides a visual summary of the construction of the regional frontiers for the unconditional (Figure 1a) and the spatially-conditioned DEA (Figure 1b) approaches, representing labour productivity against capital intensity for the case of different regional economies. We will focus on region A (denoted as *Reg A*). Assume the existence of geographical externalities amongst region A and their neighboring regions (denoted as *Reg NA$_1$* and *Reg NA$_2$*). The presence of these geographical externalities would imply changes in the shape of the production frontier, changing regional technical efficiency in our target region -*Reg A*-. Effectively, we would expect spatial dependence amongst the efficiencies of *Reg A*, and its neighbors *Reg NA$_1$* and *Reg NA$_2$*, and different behaviors of the spatial factors could produce distinct shapes of the efficient frontier for *Reg A*. In the standard approach (unconditional DEA, Figure 1a), the measure of inefficiency for region A, represented by the vertical distance from *Reg A* to the production frontier, does not consider the presence of spatial externalities derived from the existence of geographical spillovers amongst region A and their neighboring regions A$_1$ and A$_2$. However, in the SpDEA approach (spatially-conditioned case, Figure 1b), apart from the unconditional frontier, there exists a specific frontier for the case of region A that depends on the level of a set of spatial factors related to the location of this region (contextual variables that influence the frontier level of region A). In Figure 1b, the measure of inefficiency for region A would be considering the presence of geographical externalities produced from its neighboring regions. Now, the vertical distance from *Reg A* to the spatially-conditioned frontier is showing a different quantification of the inefficiency than that obtained for the unconditional case. What the figure reveals is that, under the presence of geographical externalities, the estimation of the regional inefficiencies with the standard DEA for region A are down-biased. However, while a standard DEA productivity analysis might yield results that would not reflect the impact of geographical externalities, our SpDEA approach would try to reduce the potential for such bias.

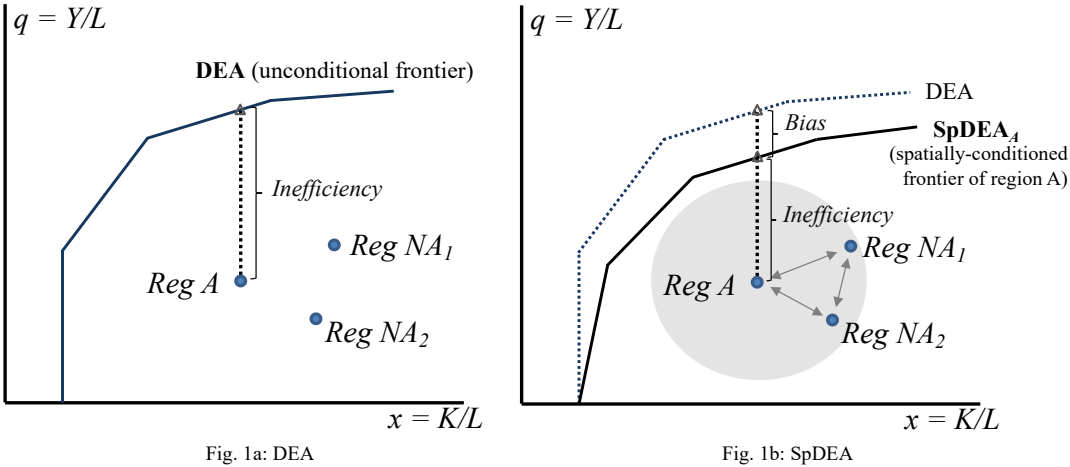

Figure 1. Unconditional and spatially-conditioned DEA regional frontiers.

Thus, we may postulate a consensus on the thesis that from a hypothetical regional empirical application using the DEA and the SpDEA models, the expectation would be an increase in the scores. Less agreement can be claimed for the role played by the existing spatial dependence on the shape of the distribution of the estimated technical efficiencies, and this is where one of the contributions of our method lies.

## 3. Empirical application

In this section, region-level technical efficiency in the EU-28 over the 2000-2014 period will be analyzed applying the standard DEA and the new SpDEA model proposed in the previous section in order to account for the presence of heterogeneous conditions in the form of geographical externalities.

**Data**

With respect to the DMUs used in our application, the empirical analysis focuses on 263 NUTS-2 regions in the (before 2020) 28 European Union countries, excluding the overseas territories of Finland, France, Portugal and Spain. The data used in the empirical application were taken from the Cambridge Econometrics' European Regional Database (ERD) 2016 release that contains complete yearly information for the period 1990-2014 at the regional NUTS-2 classification of the European Union. The primary source of the ERD is the Eurostat's REGIO database, supplemented with the European Commission's AMECO database; the 2016 release of ERD uses the NUTS 2010 regional classification.

From the ERD, the following variables were calculated or estimated:

- Regional output ($Y$), measured as gross value added -GVA- in each region in constant 2005 purchasing power standards -PPS- terms. The original GVA at constant prices time series (measured in €2005m) were adjusted for price differences across countries and over the time with country-specific PPSs.

- Regional labor ($L$), measured as total employment in each region in 000s of people.

- Regional PPS levels of GVA were divided by the total number of workers, and thus a series of real GVA per worker ($q$) were calculated for each region.

- Gross fixed capital formation ($I$), measured in €2005m.

- To obtain estimations of regional capital stocks ($K$), the Perpetual Inventory Method (PIM) was employed using yearly real gross fixed capital formation ($I$) series based on the equation $K_{it} = I_{it} + (1-\delta)K_{i,t-1}$. The investment series for the period 1990-2000 were accumulated to obtain an initial estimate of regional level of real net capital stock in the year 2000, $K_{i0} = K_{i,2000}$. Thereafter, all the stocks in this year were benchmarked to the aggregate estimates of the national capital stocks in the year 2000 (measured in €2005m) reported in Penn World Tables 8.1 (Feenstra et al., 2015) using the share of each region in the national capital stock in the year 2000 obtained with the 1990-2000 investment-accumulated regional capital stock estimations. The PIM formula was then applied to calculate the capital stock estimates for the period 2001-2014, using a depreciation rate of 5%.

   - Regional series of per worker capital ($x$) were obtained dividing regional levels of physical capital stocks by the total number of workers of each region.

**Results**[4]

In this subsection the empirical results of the analysis for the case of the 263 EU-28 NUTS-2 regions will be presented and discussed.

*Comparing DEA and SpDEA results* [5]

First, we will compare the regional technical efficiency scores estimated with both the baseline DEA model (unconditional analysis) and the SpDEA model (conditional specification) proposed in this paper.

Standard DEA scores are estimated using (2) and, drawing on (6), the SpDEA technical efficiency of an observed region ($x_{it} = K_{it}/L_{it}, q_{it} = Y_{it}/L_{it}$) facing external conditions ($z_{1i} = \mathbf{w}'_i \mathbf{q}, z_{2i} = \mathbf{w}'_i \mathbf{x}$) is given by:

$$\hat{\lambda}_{SpDEA}(x_i, q_i | \mathbf{w}'_i \mathbf{q}, \mathbf{w}'_i \mathbf{x}) = sup\{\lambda > 0 \mid \hat{S}_{Q|X,Z_1 Z_2}(\lambda q_i | X \leq x_i, Z_1 = \mathbf{w}'_i \mathbf{q}, Z_2 = \mathbf{w}'_i \mathbf{x}) > 0\} \quad (9)$$

---

[4] The R codes used in the calculations are available from the corresponding author.

[5] We are conscious that there are two issues, noisy data and endogeneity problems, not addressed in our estimation of regional efficiencies. A robust order-*m* approach could be used to obtain estimates more resistant to outliers and extreme values (Cazals et al., 2002), and the endogeneity issue could be addressed by using the procedure proposed in Simar et al. (2016). These extensions will be the objective of future consideration.

Table 1 provides summary statistics of the efficiency estimates of both models in the years 2000, 2007 and 2014. Complementarily, Figure 2 displays violin plots of the DEA and SpDEA efficiency scores for further interpretation of the results.

Table 1. Descriptive statistics on DEA and SpDEA efficiency scores.

|  | 2000 | | 2007 | | 2014 | |
|---|---|---|---|---|---|---|
|  | DEA | SpDEA | DEA | SpDEA | DEA | SpDEA |
| Mean | 0.424 | 0.586 | 0.464 | 0.584 | 0.454 | 0.576 |
| Std. Dev. | 0.150 | 0.223 | 0.130 | 0.144 | 0.117 | 0.130 |
| Min. | 0.123 | 0.123 | 0.219 | 0.219 | 0.236 | 0.245 |
| Median | 0.431 | 0.627 | 0.438 | 0.607 | 0.4235 | 0.5850 |
| Max. | 1 | 1 | 1 | 1 | 1 | 1 |
| Pearson's correlation (p-value) | 0.823 (0.000) | | 0.653 (0.000) | | 0.537 (0.000) | |
| Spearman's rank correlation (p-value) | 0.835 (0.000) | | 0.684 (0.000) | | 0.493 (0.000) | |

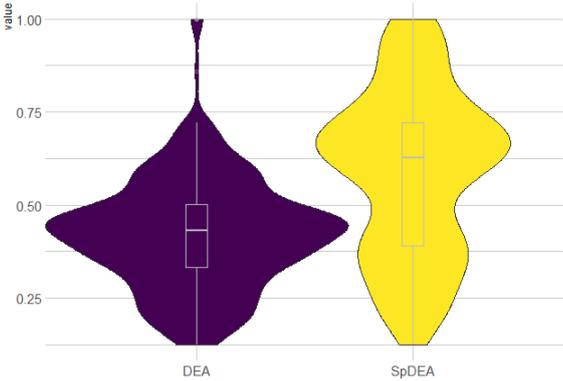

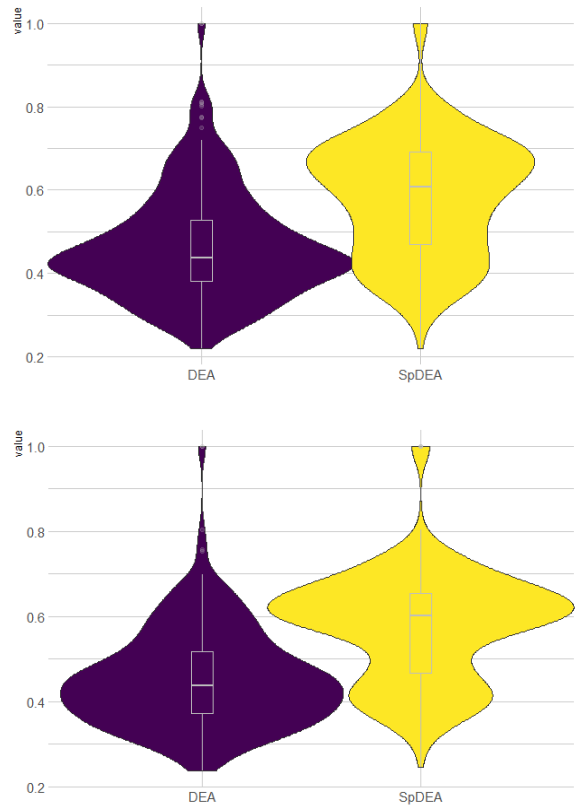

Figure 2. Violin plots of DEA and SpDEA efficiency scores (years 2000, 2007 and 2014, by rows).

Starting from the descriptive statistics, the mean of both unconditional and conditional efficiencies increases or remains approximately stable from 2000 to 2007 (from 0.424 to 0.464 in the DEA case, and from 0.586 to 0.584 in the SpDEA one) but deceases from 2007 to 2014, although the changes are modest (from 0.464 to 0.454, and from 0.584 to 0.576, respectively). Moreover, comparing the averages of regional efficiencies in every year, it can be seen that SpDEA estimates increase by 38.2%, 25.9% and 26.9%, respectively, with respect to the DEA case. These results provide evidence that a substantial part of regional technical inefficiencies in the European Union is related to interregional spillovers in the production process.

On the other hand, the standard deviation of both DEA and SpDEA efficiency scores provide straightforward indicators of sigma-convergence (narrowing of the distributions) for European regions over the period 2000-2014. Table 1 shows that the standard deviation of DEA (SpDEA) efficiency scores declined from 0.150 (0.223) in the year 2000 to 0.117 (0.130) in the year 2014. Complementarily, if the correlations of the efficiency scores for the DEA and SpDEA models are compared, it can be appreciated that Pearson's and Spearman's rank coefficients vary from about 0.823 and 0.835 in 2000 to 0.537 and 0.493, respectively, in 2014. Hence, although there has been a reduction in the dispersion of the efficiency scores (sigma-convergence), the relationship between the two scores variables has decreased during this period of time.

In summary, the above descriptive results show that the impact of neighbors on a region's level of efficiency can be considerable, indicating the importance of incorporating spatial effects into the DEA methodology so widely used in efficiency analyses. From a graphical point of view, Figure 2 complements the previous comment. Violin plots allow a deeper comparison of the scores provided by the DEA and SpDEA models by showing simultaneously the differences in location, scale, symmetry, extreme observations, and distributional characteristics of such scores values. The basic descriptive information provided by the box plots corroborates the diagnosis presented in Table 1. However, the density traces provide new information, indicating the shape of the distributions for the technical efficiency estimates: the scores show bimodal distributions –twin peaks- in the case of the SpDEA model, especially in the years 2000 and 2014, pointing to the presence of two clusters ('efficiency clubs') of regional scores. This finding is very relevant since, to the best of our knowledge, no other study has detected the emergence of bimodal regional technical efficiency in European regions. This statistical property of multimodality in the distribution of the SpDEA scores is corroborated in Table 2, where we provide the results of the Excess-Mass (EM) tests for multimodality (Ameijeiras-Alonso et al., 2019) of DEA and SpDEA efficiency scores. From the results reported in Table 2, we have determined the number of modes in the different distributions of DEA and SpDEA efficiency scores. Note that the number of modes in the distributions of DEA efficiency scores is one for all the years, while the distributions of the SpDEA estimates present two modes in the years 2000 and 2014, showing the polarization of the European regional technical efficiency. This result highlights, again, the bias of the baseline DEA in the construction of the production-frontier.

Table 2. Tests for multimodality of DEA and SpDEA efficiency scores.

| Year | DEA | SpDEA |
|---|---|---|
| 2000 | EM = 0.027<br>p-value = 0.838<br>Est. loc. = 0.427 | EM = 0.073<br>p-value < 0.000<br>Est. loc. = 0.369; 0.667 |
| 2007 | EM = 0.031<br>p-value = 0.546<br>Est. loc. = 0.423 | EM = 0.033<br>p-value = 0.558<br>Est. loc.= 0.653 |
| 2014 | EM = 0.031<br>p-value = 0.600<br>Est. loc. = 0.423 | EM = 0.069<br>p-value < 0.000<br>Est. loc. = 0.410; 0.613 |

NOTE: Excess-Mass -EM- test [$H_0$: One Mode; $H_1$: More Than One Mode], and estimated location of modes -Est. loc-. Bootstrapped p-values are used, with 5000 bootstrap replications.

*Marginal local impact of z's on the regions' efficiency scores*

To assess the direction of the influence of the spatial external variables ( $z_{1i} = \boldsymbol{w}_i'\boldsymbol{q}$ and $z_{2i} = \boldsymbol{w}_i'\boldsymbol{x}$ ) on the production process, we investigate the partial regression plots of the ratios of conditional over unconditional efficiency estimates, $\hat{R}$, with respect to these conditional variables. Figure 3 displays the results of these graphs for the years 2000, 2007, and 2014.

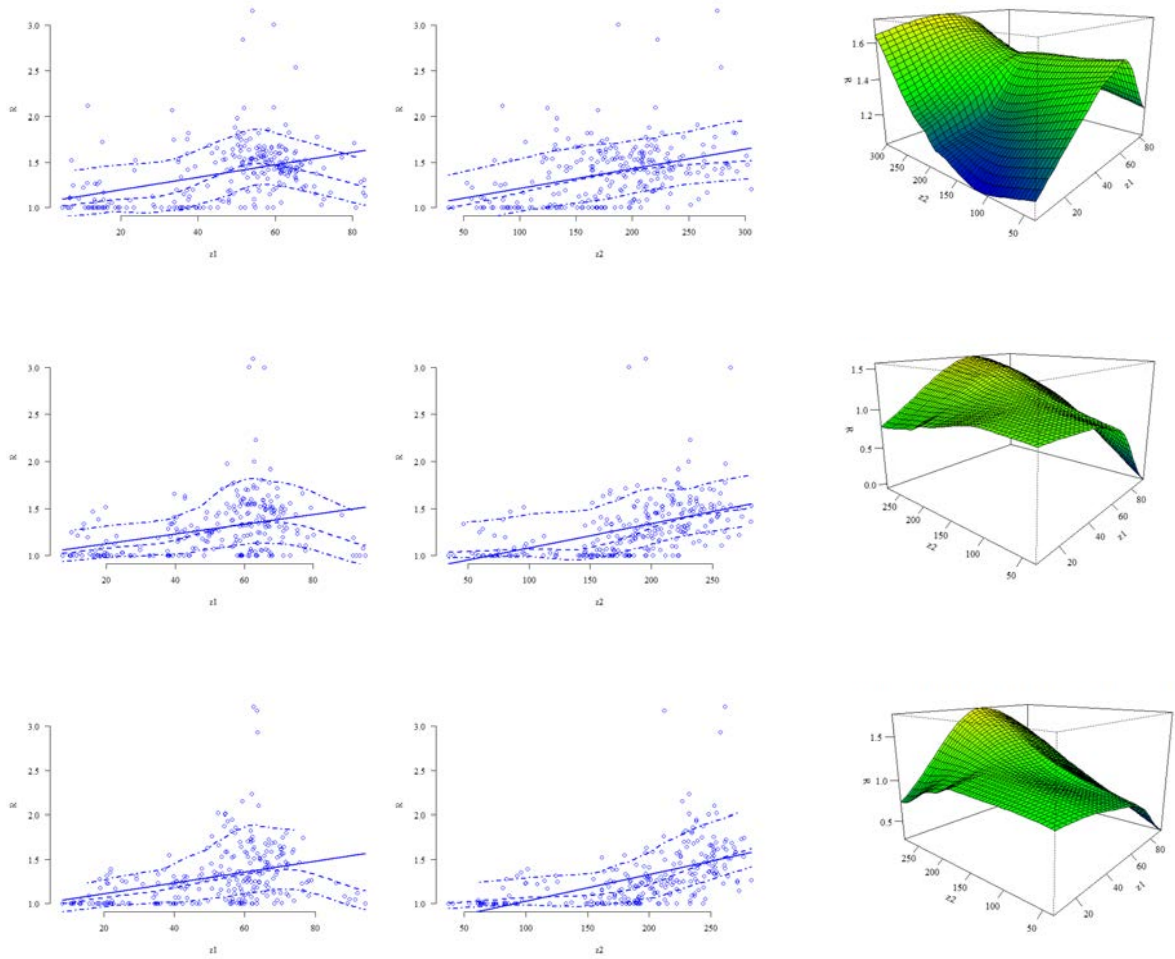

Figure 3. Local impact of spatial external variables ($z_{1i} = w'_i q, z_{2i} = w'_i x$) on the production process (years 2000, 2007 and 2014, by rows).

Focusing on the effect of $z_1$ (spatial lag of the output per worker ratio) on the shift of the frontier, we can see an approximately inverted 'U' shape form in the local regression lines. Then, we might have some shift of the efficient boundary when the real GVA per worker ratio of the neighbors increases, but after a certain threshold in the level of the output per labor variable, a decreasing effect is revealed. Hence, regional labor productivity appears to play a significant role in accelerating regional technological change (shifts in the production frontier). The reported threshold effect seems to confirm the findings of Ramajo et al. (2017) for 120 regions in EU-9 over the 1995-2007 period, where the GVA spatial lag variable also has an inverted U-shaped non-linear impact on the performance of regions, interpreting this finding as a differential effect of interregional spillovers depending on the size of the neighboring economies.

On the other hand, there is a more linear shape for the effect of $z_2$ (spatial lag of the per worker capital input) on the regions' efficient boundary in the three years considered. When the capital-labor ratio grows, shifts of the EU regional frontier occur, the *K/L* variable acting as an accelerator of the technical change. This evidence points to the favorable effect of the level of capital intensity on the production process via technology change.

Additionally, in Table 3, we test the statistical significance of each contextual variable using the kernel-weighted local linear least squares non-parametric method and the significance test discussed in section 2. As can be seen through the corresponding *p*-values, the two spatial external variables, $z_1$ and $z_2$, are statistically significant in years 2007 and 2014, but only $z_1$ is weakly significant in the year 2000. Therefore, it seems that the environmental variables used in this work had a lesser impact on the input-output space at the beginning of the period under analysis.

Table 3. Effects of spatial external variables in the SpDEA model.

|  | 2000 | 2007 | 2014 |
|---|---|---|---|
|  | p-value | p-value | p-value |
| $z_{1i} = \boldsymbol{w}_i'\boldsymbol{q}$ | 0.09* | <0.001*** | <0.001*** |
| $z_{2i} = \boldsymbol{w}_i'\boldsymbol{x}$ | 0.41 | <0.001*** | <0.001*** |

NOTE: Kernel regression significance tests (type I test with IID bootstrap: 1000 replications).

*Mapping and exploratory spatial analysis of the SpDEA efficiency scores*

Taking into account the above results with respect to the statistical significance of the spatial contextual variables, the technical efficiency estimates derived from the SpDEA model are mapped and described. In the first place, the first row of Figure 4 shows the maps of the regional SpDEA efficiency scores in 2000, 2007 and 2014, and in the second row the associated Moran's scatterplots and *I* statistics are shown.

Looking at the choropleths, there is clear positive spatial dependence between the yearly regional efficiency scores, this spatial autocorrelation being confirmed by the scatterplots and the Moran's *I* statistics (0.763 in 2000, 0.616 in 2007, and 0.541 in 2014, with pseudo *p*-values less than 0.001 in all cases).[6]

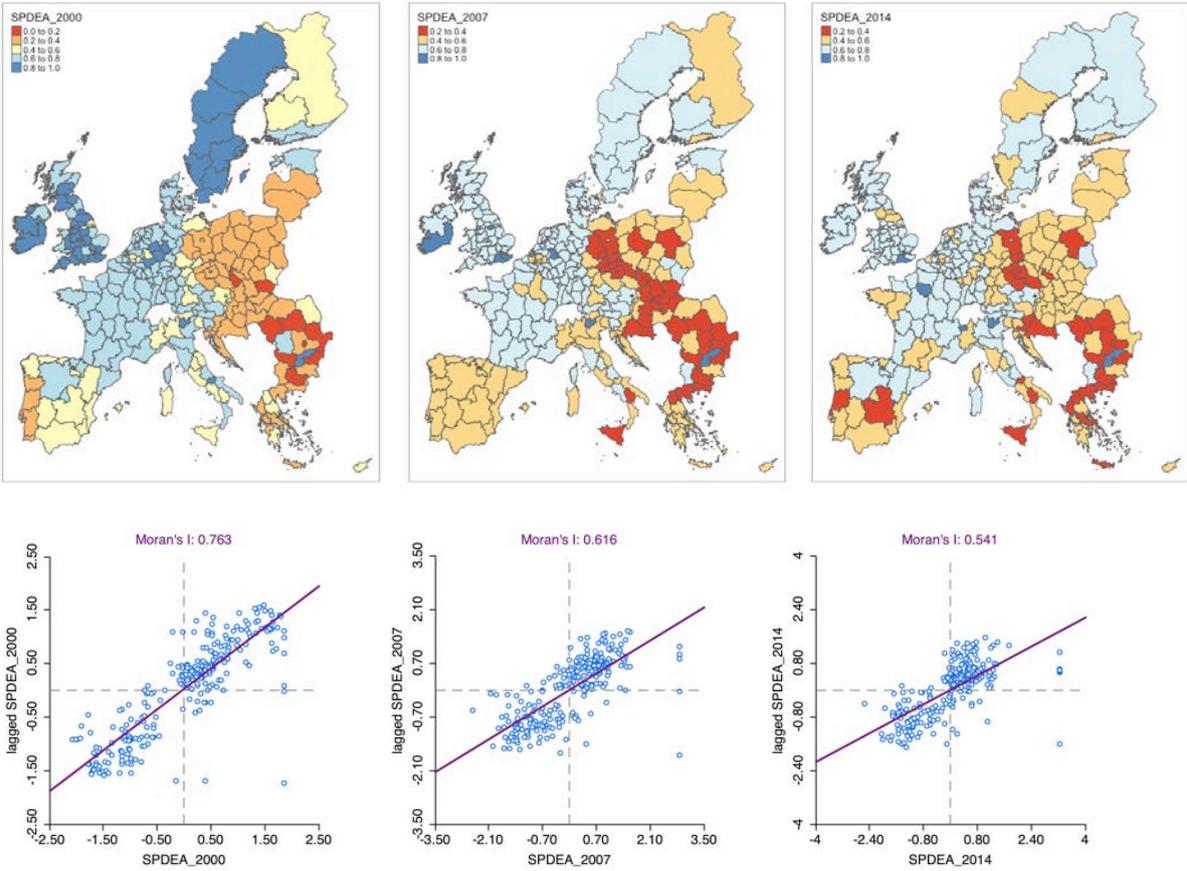

---

[6] It is important to highlight that the temporal evolution of the three Moran's scatterplots shows a final pattern (year 2014) where almost all the regions are located in the first (high-high -*HH*- efficiency scores) and third (low-low -*LL*- efficiency scores) quadrants. This pattern was corroborated by the corresponding LISA maps (not presented here due to space restrictions, but available from the authors upon request), where the existence of two spatial clusters for the EU-28 regions can be identified: the first cluster (the *HH* zone) is characterized by highly efficient regions surrounded by regions with similar scores, while the second cluster (the *LL* zone) includes more inefficient regions. Geographically, the first cluster is located in the United Kingdom and in central and northwestern Europe ('*core*'), while the second group includes the regions in eastern Europe and the most southern continental regions of Spain and Portugal ('*periphery*').

Figure 4. Choropleths and Moran's scatterplots of SpDEA efficiency scores (years 2000, 2007 and 2014, by columns).

Finally, the means of the SpDEA regional efficiency scores for each of the EU-28 countries in years 2000, 2007 and 2014, and the average over the full period 2000-2014 would provide additional analyses (these results are available under request by the authors).

*Discussion of the results and policy implications*

Our results identify geographical externalities as partially responsible for changes in technical efficiencies in the European regions. First, regional technological change can be induced by the output per worker ratio of the neighboring regions through the different microfoundations of agglomeration economies that could generate production cost savings. Secondly, the proximity or density of high capital per worker inducing technology change is another mechanism through which the regional technical efficiencies could change.

Also our estimates of regional technical efficiencies reveal that, over the analyzed period, the efficiency scores have bimodal distributions. This is a new finding for the European regional case that could be related to the detected multimodal distribution in European regional productivity. For example, the kernel distributions of the regional efficiencies estimated by Rogge (2019) across NUTS-2 regions in the EU for the years 2000 and 2011 demonstrate that the distribution of regional productivity was multimodal, but do not detect the existence of multimodality in the distribution of efficiency scores. This negative detection could be the result of the a-spatial nonparametric production frontier technique used by Rogge (2019) to estimate the scores.

Additionally, in general, the regional distribution of labor productivity in Europe has shown the existence of two clubs of convergence (Ramajo et al., 2008). These clubs maintain internal relationships characterized by a geographical pattern (see, for example, Basile, 2009, or Escribá-Pérez and Murgui-García, 2018). From our results, it could be hypothesized that this polarization in regional productivity could be driven primarily by a bimodal technical efficiency distribution, This would imply that the two clubs of European regions in terms of regional convergence have their counterfactual part into two categories of European regions that are showing different technical efficiencies.

To try to explain the polarization found in our estimated SpDEA scores, we will carry out a complementary descriptive analysis that tries to correlate the estimated scores for the year 2014 with the European Regional Competitiveness Index (RCI) proposed by the European Commission to measure the major factors of competitiveness of the regions within the EU.[7]

The RCI is a measure of the performance of the NUTS-2 regions within the EU, and its 11 dimensions of competitiveness -pillars- (each one capturing concepts related to regional productivity and long-term economic growth) are grouped into three sub-indices: basic (BSI), including (1) institutions, (2) macroeconomic stability, (3) infrastructure, (4) health, and (5) basic education; efficiency (ESI), including (6) higher education, training and lifelong learning, (7) labor market efficiency, and (8) market size); and innovation (ISI), including (9) technological readiness, (10) business sophistication, and (11) innovation.

The Pearson's correlation that links our 2014 SpDEA efficiency scores with the RCI variable yields a statistically significant value of 0.557 (p-value < 0.001). This is a sign that our measure of technical efficiency is related in part to the general competitiveness of the European regional economies. Moreover, the correlations for the case of the three sub-indices are relatively similar: basic -BSI- (0.540), efficiency -ESI- (0.558) and innovation -ISI- (0.503).[8]

---

[7] The most recent version of the RCI is the one published in 2019 (Annoni and Dijkstra, 2019). The RCI 2016 is based on more than 70 comparable indicators which are mostly referred to the period 2013–2014, our year of reference (2014) for the ongoing descriptive correlation analysis. The RCI indices were launched in 2010 and published every three years; therefore, there is no information for the RCIs that permits comparison with the SpDEA scores of the years 2000 and 2007.

These correlations indicate that the SpDEA efficiency scores retain, however, the information contained in the RCI and its main dimensions of competitiveness as measures of the performance of the NUTS-2 regions within the EU, whereas correlations for the traditional DEA efficiency scores present lower correlations (RCI = 0.434; BSI =0.332; ESI = 0.467 and ISI = 0.392). These coefficients also support the claim that, unlike the SpDEA efficiency scores, which are sensitive to the regional spatial dependence, the use of the traditional DEA efficiency

---

[8] It is necessary to emphasize that, considering the 11 pillar scores of the RCI, the highest correlations within the group "Basic" corresponds to "Institutions" (0.524) and "Infrastructure" (0.491). Institutions is an indicator of the quality of government that reflects the perceptions of the citizens with respect to corruption, quality and impartial allocation in their public-sector services. The important role of institutions for economic performance has been highlighted recently by Agostino et al. (2020) and Rodríguez-Pose and Ganau (2021) when analysing the determinants of regional labor productivity trajectories in Europe. Besides, the contribution of infrastructures to the productivity across European regions has been studied by numerous studies (see, for example, Del Bo and Florio, 2012, and Basile, 2009).

Within the group of "Efficiency," the most important components in terms of correlation are "Labour Market Eficciency" (0.503) and "Market Size" (0.560). With respect to the market size pillar, this indicator embraces characteristics such as regional GDP and population. The relation between market size and productivity has been recognized by different authors (for example, Bernhardt, 1981, and Epifani and Gancia, 2006). Other authors have examined the influence of labour market efficiency on European productivity (Escribá-Pérez and Murgui-García, 2018; Cyrek and Fura, 2019).

Finally, within the third group, "Innovation," the highest correlations are found for "Technological Readiness" (0.483) and "Business Sophistication" (0.451). The positive influence of the innovation indicator, related to the use of information and communication technologies, has been emphasized, among others, by Whelan (2002) and Griffith et al. (2004), while authors like Antonelli et al. (2011) or Siller et al. (2021) have been focused on the relevance of knowledge and technology-intensive sectors (business sophistication).

Therefore, these partial correlations could be indicating, in part, the underlying mechanisms responsible for the existence of low or high levels of efficiency in an important group of European regions. Technical efficiency is the key way that regions can raise productivity and so, to increase their per-capita income en route to a path towards achieving regional economic convergence.

scores could produce unrealistic diagnoses about the regional performances, with subsequent wrong economic policy recommendations.

Additionally, although it has not been noted previously in the literature, the determinants of the existence of twin peaks in the efficiency of the European regions would be important indicators of the causes of the differences in productivity between the US and Europe. Indeed, European productivity slowdown relative to the United States of America has been an important focus for applied researchers. In particular, among the diverse studies, a lot of attention has been paid to the slower emergence in Europe of the so called "knowledge-based economy" (Griffith et al. 2006, Van Ark et al., 2008). The efficiencies derived from the Information and Communication Technologies within a regional economy (and their associated geographical externalities) could be a key element to stimulate the European productivity (Miller and Atkinson, 2014). Nevertheless, this could be a simplistic expression of a more complex problem: while the 'knowledge economy' is a fundamental explanation for the European productivity gap, the effects from another causes could be also operating (Ortega-Argilés, 2012). Hence, at the regional level, other factors, such as institutional qualities, market size or the regional innovative capacity may matter for regional productivity (see Pike et al., 2017).

From our empirical results, future policy challenges related to European productivity should attempt to foster efficiency through economic policies that focus on the presence of regional neighborhood effects. This implies addressing the geographical externalities of regional economic policies beyond the single region objectives they pursue. Comprehensive policies that cover the operational field of interregional spillovers could be in the design of measures ready to break the existence of dichotomies within European regional economies through the existence of twin peaks in regional productivity, regional convergence and, as shown in this paper, regional efficiency.

# 4 Summary and concluding remarks

In this paper, it is argued that the standard DEA method used to estimate efficiencies with spatial data is inappropriate when dealing with DMUs whose efficiencies are correlated with the performances of the neighboring units. In order to illustrate the relevance of our proposal, the results from the traditional DEA and the SpDEA models have been considered for the case of the analysis of 263 NUTS-2 European regions. The results obtained with the SpDEA model show that in standard DEA applications, without the appropriate spatial corrections, inefficiency is overestimated, providing evidence that a non-negligible part of regional technical efficiency in the European Union is related to interregional spillovers in the production process. This finding could challenge the appropriateness of policies that have, for example, under-allocating public resources to regions with highly significant geographical spillovers, or raise doubts about the estimation of the level of 'budgetary waste' in EU Member States' spending applying standard DEA to countries or regions' production levels of public services (Saulnier, 2020).

On the other hand, the dispersion of the spatially-conditioned scores has declined over time, showing that EU-28 regional technical efficiency has sigma-converged during the period under analysis, 2000-2014. Furthermore, the SpDEA scores show bimodal distributions - twin peaks - especially in the years 2007 and 2014, pointing to the presence of two spatial clusters (or efficiency clubs) of regional scores. Although this distributional pattern has been identified previously in the literature in terms of both European regional productivity and convergence of per capita incomes, the detection of twin peaks in the regional efficiency within EU is a new empirical result.

As general conclusion, the SpDEA model can help to explain efficiency differentials between European regions that contribute to the generation of different regional productivity growth trajectories. A future research priority should be to investigate the underlying mechanisms that are maintaining these two clubs of efficiency in the European regions. Another future development would be to extend our analysis at sectoral level, studying the efficiencies of different industries.

An implication of our results is that the technical efficiency in the European regions may be underestimated in studies that use the standard DEA model in order to try to explain the low levels of regional productivity, hiding the peaks and the troughs in the spatial distribution of regional technical efficiencies. A more high-level consequence of the spatial heterogeneity analysis carried out in this paper is that contemporaneous inter-country or multi-regional macroeconomic models need to take into account the existence of spatial spillovers between cross-units, and therefore explicitly include spatial aspects in its formulation. These spatially-based theoretical models, and their empirical counterparts, will be more useful in order to better understand issues such as global income inequality, productivity convergence between countries, or differential regional impacts of any innovation within the EU or any other geographic area, to cite just a few examples.